# The Cosmic Axion Spin Precession Experiment (CASPEr): a dark-matter search with nuclear magnetic resonance.

**Antoine Garcon**[1,2], **Deniz Aybas**[4,5], **John W. Blanchard**[2], **Gary Centers**[1,2], **Nataniel L. Figueroa**[1,2], **Peter W. Graham**[9], **Derek F. Jackson Kimball**[3], **Surjeet Rajendran**[6], **Marina Gil Sendra**[1,2], **Alexander O. Sushkov**[4], **Lutz Trahms**[8], **Tao Wang**[6], **Arne Wickenbrock**[1,2], **Teng Wu**[1,2], **and Dmitry Budker**[1,2,6,7]

[1] Johannes Gutenberg-Universität, Mainz 55128, Germany
[2] Helmholtz Institute, Mainz 55099, Germany
[3] Department of Physics, California State University East Bay, Hayward, California 94542-3084,USA
[4] Department of Physics, Boston University, Boston, Massachusetts 02215, USA
[5] Department of Electrical and Computer Engineering, Boston University, Boston, Massachusetts 02215, USA
[6] Department of Physics, University of California, Berkeley, CA 94720-7300,USA
[7] Nuclear Science Division, Lawrence Berkeley National Laboratory, Berkeley, CA 94720,USA
[8] Physikalisch Technische Bundesanstalt, Abbestrasse 2-12, 10587 Berlin, Germany
[9] Department of Physics, Stanford Institute for Theoretical Physics, Stanford University, Stanford, California 94305, USA

E-mail: garcon@uni-mainz.de

**Abstract.** The Cosmic Axion Spin Precession Experiment (CASPEr) is a nuclear magnetic resonance experiment (NMR) seeking to detect axion and axion-like particles which could make up the dark matter present in the universe. We review the predicted couplings of axions and axion-like particles with baryonic matter that enable their detection via NMR. We then describe two measurement schemes being implemented in CASPEr. The first method, presented in the original CASPEr proposal, consists of a resonant search via continuous-wave NMR spectroscopy. This method offers the highest sensitivity for frequencies ranging from a few Hz to hundreds of MHz, corresponding to masses $m_a \sim 10^{-14}$–$10^{-6}$ eV. Sub-Hz frequencies are typically difficult to probe with NMR due to the diminishing sensitivity of magnetometers in this region. To circumvent this limitation, we suggest new detection and data processing modalities. We describe a non-resonant frequency-modulation detection scheme, enabling searches from mHz to Hz frequencies ($m_a \sim 10^{-17}$–$10^{-14}$ eV), extending the detection bandwidth by three decades.

## 1. Introduction

In 1945 Edward M. Purcell measured the first radio-frequency absorption from nuclear magnetic moments in paraffin [1]. In the following months, Felix Bloch observed nuclear spin precession in water [2]. Subsequently, Bloch's nuclear magnetic resonance (NMR) techniques showed that electrons provide magnetic shielding to the nucleus. The resulting change in magnetic-resonance frequency, known as the chemical shift, provides information on the electronic environment of nuclear spins. The discovery of this phenomenon enabled NMR-based chemical analysis [3] and the field of NMR quickly grew to become a dominant tool in analytical chemistry, medicine and structural biology. NMR also remains at the forefront of fundamental physics, in fields such as materials science, precision magnetometry, quantum control [4, 5] and in searches for exotic forces. In this vein, we discuss how NMR can be applied to the detection of dark matter, in particular cold-axion dark matter.

The existence of dark matter was postulated in 1933 by Fritz Zwicky to explain the dynamics of galaxies within galaxy clusters. Zwicky discovered that the amount of visible matter in the clusters could not account for the galaxies' velocities and postulated the presence of some invisible (dark) matter [6]. Astrophysical observations now show that dark matter composes more than 80% of the matter content of the universe. A multitude of particles were introduced as possible candidates, but as of today, none of them have been detected and the nature of dark matter remains unknown. The axion particle, emerging from a solution to the strong-CP problem, is a well-motivated dark matter candidate.

The Standard Model predicts that the strong force could violate the charge conjugation parity symmetry (CP-symmetry) but this has never been observed and experimental measurements constrain strong-CP violation to an extremely low value [7]. The need for this fine tuning is known as the strong CP problem. In 1977, Roberto Peccei and Helen Quinn introduced a mechanism potentially solving the strong CP problem [8], from which the axion emerges as a bosonic particle [9, 10].

Low-mass axions and other axion-like particles (ALPs) could account for all of the dark matter density [11, 12, 13]. Indeed, axions and ALPs interact only weakly with particles of the Standard Model, making them "dark". In addition, axions and ALPs can form structures similar to the dark-matter galactic clusters [14, 15].

In addition to interacting via gravity, axions and ALPs are predicted to have a weak coupling to the electromagnetic field, enabling their conversion to photons via the inverse-Primakoff effect [16, 17]. Past and current axion searches largely focus on detecting photons produced by this coupling. These searches include helioscopes, detectors pointed at the Sun, such as the "CERN Axion Solar Telescope" (CAST [18]). Helioscope detection could happen when axions produced in the Sun are converted back to photons in a strong laboratory magnetic field [19].

"Light-Shining-Through-Walls" experiments, such as the "Any Light Particle Search" (ALPS [20]), do not rely on astrophysical-axion sources. These experiments seek to convert laser-sourced photons into axions in a strong magnetic field. Subsequently, axions travel through a wall and are converted back to photons via the same mechanism.

Other searches include haloscopes, aimed at detecting local axions in the Milky Way's dark-matter halo. These experiments include microwave cavity-enhancement methods, such as the "Axion Dark Matter Experiment" (ADMX [21, 22]), which could convert local axions to photons in a high-Q cavity. More information on the wide array of experimental searches for axion and ALP dark matter can be found in Ref. [23].

The possibility of direct ALP detection via their couplings to nucleons and gluons has been recently proposed [24, 25]. These couplings give rise to oscillatory pseudo-magnetic interactions with the dark matter axion/ALP field. This dark matter field oscillates at the axion/ALP Compton frequency, which is proportional to the axion/ALP mass. The Cosmic Axion Spin Precession Experiment (CASPEr [24]) is a haloscope, seeking to detect the NMR signal induced by these couplings. The CASPEr collaboration is composed of two main groups, each searching for axions and ALPs via different couplings: CASPEr-Wind is sensitive to the pseudo-magnetic coupling of ALPs to nucleons and CASPEr-Electric is sensitive to the axion-gluon coupling. The two experiments rely on different couplings but are otherwise similar, in the sense that they both measure axion/ALP-induced nuclear-spin precession.

In the following section, we explain how the ALP-nucleon and axion-gluon couplings could induce an NMR signal. Subsequent sections focus on describing two NMR measurement schemes implemented in both CASPEr-Wind and CASPEr-Electric. The first method, presented in the original CASPEr proposal [24], consists of a resonant search via continuous-wave NMR spectroscopy (CW-NMR). This method offers the highest sensitivity for frequencies ranging from a few Hz to hundreds of MHz. Sub-Hz frequencies are typically difficult to probe with NMR due to the diminishing sensitivity of magnetometers in this region. We present a non-resonant frequency-modulation scheme that may circumvent this limitation.

## 2. Axion- and ALP-induced nuclear spin precession

### 2.1. ALP-nucleon coupling - CASPEr-Wind

CASPEr-Wind is a haloscope searching for ALPs in the Milky Way's dark-matter halo via their pseudo-magnetic coupling to nucleons, referred as the ALP-nucleon coupling. As the Earth moves through the galactic ALPs, this coupling gives rise to an interaction between the nuclear spins and the spatial gradient of the scalar ALP field [25]. The Hamiltonian of the interaction written in Natural Units takes the form:

$$H_{aNN} = g_{aNN} \sqrt{2\rho_{DM}} \cos(m_a t) \vec{v}.\vec{\sigma}_N, \tag{1}$$

where $\vec{\sigma}_N$ is the nuclear-spin operator, $v \sim 10^{-3}$ is the velocity of the Earth relative to the galactic ALPs, $\rho_{DM} \sim 0.4$ GeV/cm$^3$ is the local dark-matter density [26] and $g_{aNN}$ is the coupling strength in GeV$^{-1}$. The ALP mass, $m_a$, usually given in electron-volts, can also be expressed in units of frequency, more relevant for an NMR discussion. The Compton frequency associated to the axion and ALP mass is given by: $\omega_a = m_a c^2/\hbar$, where $c$ is the speed of light in vacuum and $\hbar$ is the reduced Plank constant. For the rest of the discussion, we set $\hbar = c = 1$.

The coupling in Eq. (1) is the inner product of an oscillating vector field with the nuclear-spin operator. Therefore Eq. (1) can be rewritten as an interaction between spins and an oscillating pseudo-magnetic field:

$$H_{aNN} = \gamma \vec{B}_{ALP}.\vec{\sigma}_N, \tag{2}$$

where $\gamma$ is the gyromagnetic ratio of the nuclear spin and we have identified the ALP-induced pseudo-magnetic field known as the "ALP wind":

$$\vec{B}_{ALP} = g_{aNN} \frac{\sqrt{2\rho_{DM}}}{\gamma} \cos(\omega_a t) \vec{v}. \tag{3}$$

Equation (3) can be understood as follows: as nuclear spins move with velocity $\vec{v}$ through the galactic dark-matter halo, they behave as if they were in an oscillating-magnetic field $\vec{B}_{ALP}$ of frequency $\omega_a$, oriented along $\vec{v}$. As $\rho_{DM}$ and $\vec{v}$ are determined by astrophysical observations, the only free parameters are the ALP frequency (or equivalently, the ALP mass) and the coupling constant, which define the two-dimensional parameter space of the ALP-nucleon coupling shown in Fig. 1. Thus the measured amplitude of $\vec{B}_{ALP}$ probes the value of $g_{aNN}$. Considering the coupling constant range of interest in Fig. 1 ($g_{aNN} \sim 10^{-3}$–$10^{-23}$ GeV$^{-1}$) and the $^{129}$Xe nuclear gyromagnetic ratio ($\gamma \sim 11.777$ MHz/T) yields an ALP-wind amplitude spanning $|\vec{B}_{ALP}| \sim 10^{-10}$–$10^{-30}$ T. In order for an experiment targeting the ALP wind to surpass existing astrophysical and laboratory constraints on $g_{aNN}$, the experiment must be sensitive to ultralow magnetic fields.

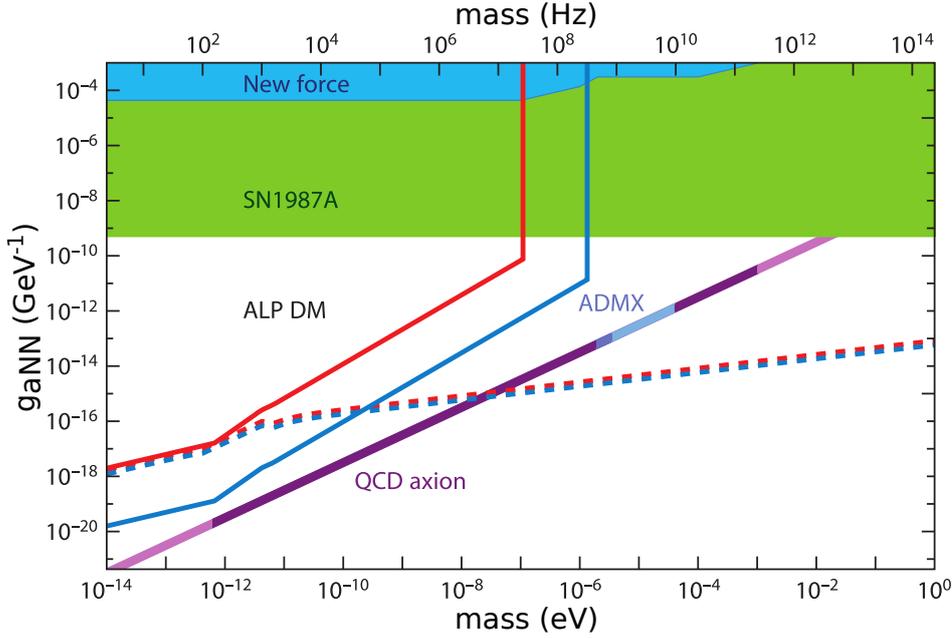

Figure 1: ALP-nucleon coupling parameter space: coupling strength $g_{aNN}$ versus ALP mass $m_a$. The purple line represents the mass-coupling parameter space corresponding to the QCD axion proposed to solve the strong CP problem [25]. The darker purple region of the line shows where the QCD axion could be all of the dark matter. The red line is the projected sensitivity of CASPEr-Wind using hyperpolarized $^{129}$Xe. The blue line is the sensitivity using hyperpolarized $^3$He during a future upgrade of the experiment. The dashed lines are the limits from magnetization noise for $^{129}$Xe (red) and $^3$He (blue). The ADMX region shows the mass range already excluded (dark blue) or that will be covered (light blue) by ADMX (probing the axion-photon coupling). The green region is excluded by observations on Supernovae SN1987A [27, 28]. The blue region is excluded by searches for new spin-dependent forces. Figure adapted with permission from Ref. [29].

The oscillating nature of the ALP wind suggests a magnetic-resonance-based detection method. In the following discussion we briefly explain why NMR techniques, specifically continuous-wave NMR, are particularly well-suited for this application.

Consider a collection of nuclear spins with gyromagnetic ratio $\gamma$, immersed in a static magnetic field $\vec{B}_0$ (the leading field; see Fig. 2-a), oriented along the z-axis. The spins orient along the leading field to produce a bulk magnetization along the z-axis. We now introduce an oscillating-magnetic field $\vec{B}_{xy}(t)$ oriented in the transverse xy-plane. If the magnitude of the leading field is such that the Larmor frequency $\gamma|\vec{B}_0|$ is equal to the oscillating field frequency, a resonance occurs. The magnetization responds by building up a transverse component, $\vec{M}_{xy}$.

Subsequently, the transverse magnetization undergoes a precession about $\vec{B}_0$, at the Larmor frequency. This oscillation creates a time-varying magnetic field that can be picked-up via magnetometers, producing the NMR signal. The spectrum exhibits a Lorentzian-shaped peak at the Larmor frequency. The search for the resonance is done by varying the magnitude of the leading field and monitoring the transverse magnetization.

This protocol is the continuous-wave NMR experiment introduced by Bloch's first nuclear induction experiment [2]. The term continuous-wave refers to the fact that the oscillating field is continuously applied to the sample. Frequencies are probed sequentially by varying the leading-field magnitude or the oscillating-field frequency.

The fact that the ALP wind has an unknown frequency and cannot be "switched off" suggests a similar experimental scheme. CASPEr-Wind is effectively a CW-NMR experiment in which the transverse component of $\vec{B}_{\text{ALP}}$ relative to $\vec{B}_0$ is analogous to the oscillating transverse field: $\vec{B}_{\text{ALP,xy}} \longleftrightarrow \vec{B}_{\text{xy}}$. The resonance is reached when the Larmor frequency is equal to the ALP frequency ($\gamma|\vec{B}_0| = \omega_a$). The experimental scheme is represented in Fig. 2-a.

This ALP-induced NMR signal is characterized with two relevant coherence times defining the linewidth of the resonance. The ALP wind oscillates with a temporal coherence inversely proportional to its frequency: $\tau_{\text{ALP}} \sim 10^6/\omega_a$ [25]. This effect is modelled by assuming that $\vec{B}_{\text{ALP}}(t)$ in Eq. (3) acquires a random phase after each time interval $\tau_{\text{ALP}}$ [24]. In addition, the transverse magnetization decoheres and decays exponentially with a characteristic time $T_2$ [30]. For the sake of an NMR-based discussion, we will now assume that the signal coherence time is limited by $\tau = \min\{T_2, \tau_{\text{ALP}}\}$. As a result the expected linewidth of the resonance becomes $\delta\nu = 1/\pi\tau$ [30] ‡.

The allowed values of the ALP mass span many orders of magnitude, yielding a large frequency bandwidth to explore (see Fig. 1). Conveniently, the NMR techniques used in CASPEr-Wind are broadly tunable, with the upper bound of the scanned region limited by the achievable magnetic-field strength. In addition, in order to avoid signal broadening that would reduce overall sensitivity, $\vec{B}_0$ must remain spatially homogeneous over the sample region. These requirements are readily met by superconducting NMR magnets up to fields of about 20 T. Such capabilities enable the detection of ALPs with corresponding frequencies ranging from a few Hz to hundreds of MHz ($m_a \sim 10^{-14}$–$10^{-6}$ eV). As such, CASPEr-Wind is a broadband search for light ALP dark matter and is complementary to many other experiments typically looking at higher mass ranges (e.g ADMX searches for axions of mass $m_a \geq 10^{-6}$ eV [23]).

## 2.2. Axion-gluon coupling - CASPEr-Electric

The second CASPEr experiment, CASPEr-Electric, relies on the coupling between CP-solving axions and gluons. This coupling induces an oscillating nucleon electric dipole moment (EDM) [33]:

$$d_n = g_d \frac{\sqrt{2\rho_{\text{DM}}}}{\omega_a} \cos(\omega_a t), \qquad (4)$$

where $g_d$ is the strength of the axion-gluon coupling in GeV$^{-2}$. As in the case of CASPEr-Wind, the only two free parameters are $g_d$ and $\omega_a$, giving rise to another two-dimensional parameter space to explore.

The major experimental difference between CASPEr-Electric and CASPEr-Wind is that CASPEr-Electric makes use of a static electric field applied perpendicularly to the leading magnetic field. As in CASPEr-Wind, $|\vec{B}_0|$ is tuned to scan for resonance. If the resonance condition is met, the axion-induced EDM oscillates at the Larmor frequency. The interaction between the EDM and the static electric field causes spins to rotate away from the direction of $\vec{B}_0$. This produces a non-zero oscillating transverse magnetization, $\vec{M}_{xy}(t)$. Subsequently, $\vec{M}_{xy}(t)$ undergoes precession about $\vec{B}_0$ and induces the NMR signal.

CASPEr-Electric and CASPEr-Wind detection schemes are similar in the sense that the effects of both, an oscillating EDM in the presence of a static transverse electric field and an oscillating ALP wind, are analogous to weak magnetic fields, oscillating at the axion or ALP frequencies. Both CASPEr-Wind and CASPEr-Electric can be described as CW-NMR experiments. Although we now focus on the sensitivity of CASPEr-Wind, the discussion for CASPEr-Electric is analogous. We note that CASPEr-Wind is not only sensitive to axions and ALPs but could also detect any light particle coupling to nuclear spins, in particular hidden photons [34, 35].

‡ Recent theories suggest that the signal should not necessarily be Lorentzian-shaped but could be asymmetric, reflecting the fact that the ALPs energy cannot be smaller than $m_a c^2$ (for a stationary ALP) and is higher by $m_a v^2/2$ for a moving ALP [31]. However the discussion remains equivalent.

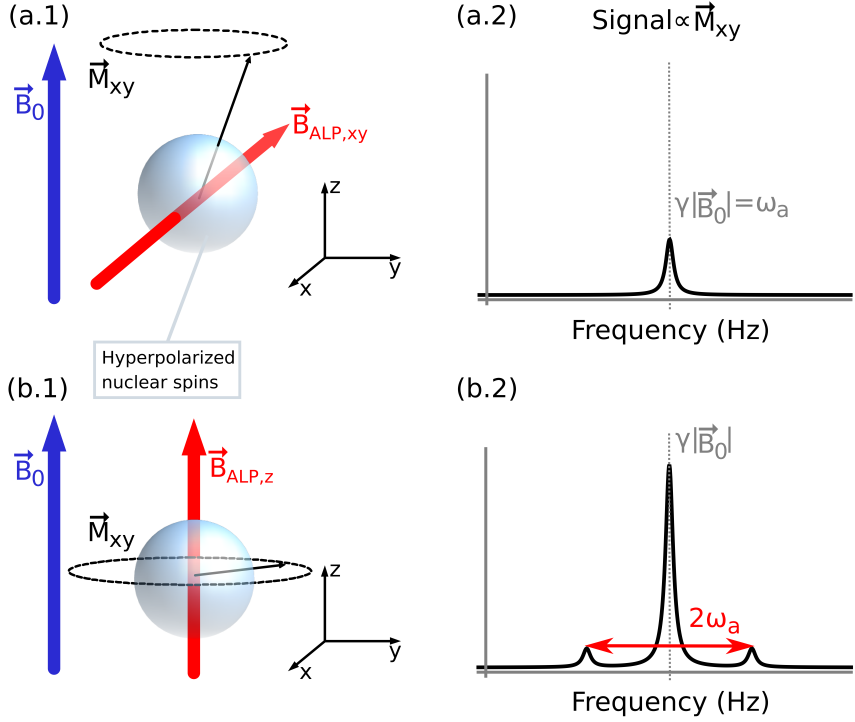

**Figure 2:** Schematic representation of the resonant (CASPEr) and sidebands (CASPEr, SILFIA [32]) experimental schemes. The hyperpolarized $^{129}$Xe sample is immersed in the leading field $\vec{B}_0$ produced by a tunable NMR magnet. The magnetometer is sensitive to the transverse magnetization $\vec{M}_{xy}$. The black arrow represents the instantaneous total magnetization. (a.1) Resonant scheme: at resonance ($\gamma|\vec{B}_0| = \omega_a$), the transverse component of the ALP wind, $\vec{B}_{ALP,xy}$, tilts the sample's magnetization which acquires a non-zero component on the xy-plane: $\vec{M}_{xy}$. $\vec{M}_{xy}$ precesses about $\vec{B}_0$ at the Larmor frequency. (a.2) The resonant signal is a low amplitude Lorentzian-shaped peak at the ALP frequency. (b.1) Sideband scheme: subsequently to a $\pi/2$ magnetic pulse, the magnetization is on the xy-plane. $\vec{M}_{xy}$ precesses at the Larmor frequency. The longitudinal component of the ALP wind, $\vec{B}_{ALP,z}$, induces modulation of the Larmor frequency. (b.2) The signal of the non-resonant scheme exhibits a carrier frequency ($\gamma|\vec{B}_0|$) and a set of sidebands at $\gamma|\vec{B}_0| \pm \omega_a$. The amplitude of the carrier frequency is large because the full magnetization is rotated by the $\pi/2$ pulse. The sidebands amplitude is expected to be small.

## 2.3. Sensitivity

Here we discuss the physical parameters affecting the sensitivity of the experiment. Later on, these results are compared to the ones obtained for a non-resonant detection scheme, aiming at probing lower frequencies. During this discussion, we consider the signal on resonance and limit the integration time to $\tau < \min\{\tau_{ALP}, T_2\}$, during which there is coherent averaging of the signal. At resonance, the transverse magnetization increases as if under the action of a low-amplitude rf-field [30]:

$$|\vec{M}_{xy}(\tau)| \propto \gamma\rho P \sin(\gamma|\vec{B}_{ALP,xy}|\tau) \quad \Rightarrow \quad |\vec{M}_{xy}(\tau)| \propto \gamma^2\rho P|\vec{B}_{ALP,xy}|\tau, \tag{5}$$

where $\rho$ is the spin density of the sample given in cm$^{-3}$ and P$\in [0,1]$ is the dimensionless polarization factor. The NMR signal can be written in terms of the transverse magnetization:§

$$|S(t)| \propto |\vec{M}_{xy}(\tau)| \quad \Rightarrow \quad |S(t)| \propto \gamma^2\rho P|\vec{B}_{ALP,xy}|\tau. \tag{6}$$

To make an estimation on the signal-amplitude threshold $A_{min}$, for which a event is detected, we use the model proposed in Ref. [24] and assume that the noise is dominated by the magnetometer white noise. The white-noise spectral density, $\sqrt{S(f)}$, usually given in fT/$\sqrt{\text{Hz}}$, is experimentally determined and depends on the frequency probed, $f$. Limiting the integration time to $\tau$, the signal is coherently averaged and remains constant while the white-noise spectral density decreases as $1/\sqrt{\tau}$ [36]. An event is detected if the signal amplitude is higher than the white noise spectral density after $\tau$, yielding $A_{min} \sim \sqrt{S(f)/\tau}$. The signal-to-noise ratio at resonance, SNR$_{RES}$, is obtained by comparing the signal amplitude from Eq. (6) to $A_{min}$:

$$\text{SNR}_{RES} := |S(\tau)|/A_{min} \tag{7}$$

$$\propto \gamma^2\rho P|\vec{B}_{ALP,xy}|\sqrt{S(f)}^{-1}\tau^{3/2}. \tag{8}$$

The SNR of the resonant signal increases as $\tau^{3/2}$. The reasons are as follows: 1) The integration takes place in a time window during which the signal stays phase coherent, thus scaling the SNR as $\tau^{1/2}$ 2) When the resonance condition is satisfied, $|\vec{M}_{xy}(t)|$ increases linearly with time. Hence, the signal is linearly amplified by $\tau$.

Because of this, the SNR greatly benefits from the use of samples with long spin coherence times $T_2$. Typical values of $T_2$ for liquid $^{129}$Xe are on the order of 10 to 1000 seconds [37, 38], making xenon an attractive sample. Liquid xenon also provides high spin density, and can be polarized above 50% via spin-exchange optical pumping methods [39], increasing the sensitivity by a factor of at least $10^5$ compared to thermal polarization (usually on the order of parts-per-million).

The choice of magnetometer determines the value of $\sqrt{S(f)}$ and is constrained by the frequency of the expected signal. In the $\omega_a/2\pi \sim 10$–$10^6$ Hz region, the best sensitivities are achieved by SQUIDs. Accounting for all experimental parameters (ALP-wind coherence time, sample geometry, density, and polarization), such SQUIDs would allow CASPEr to reach unconstrained regions of the parameters space (see Fig. 1; details are given in Ref. [24]).

Frequencies higher than 2 MHz are usually considered to be above the sensitivity cross-over between SQUIDs and inductive pick-up coils. To probe the 2–200 MHz region CASPEr will enter its phase II, in which the magnetometer is switched from a SQUID to inductive pick-up coils. Atomic magnetometers may also be used at low frequencies, in particular with zero and ultralow magnetic fields as suggested in Ref. [40].

## 3. Ultralow frequencies: sidebands detection

It turns out to be difficult to probe frequencies below ∼10 Hz with the previously described resonant scheme. Indeed, SQUIDs lose sensitivity below their characteristic "1/f-knee frequency", typically on the order of a few Hz [41]. To overcome this, one can use a non-resonant measurement protocol that could be implemented as a low-frequency extension to CASPEr. The method consists in measuring sidebands induced by modulation of the Larmor frequency and removes the need to scan for resonance. This experimental scheme, represented in Fig. 2-b, was first introduced by the "Sideband in Larmor Frequency Induced by Axions" experiment (SILFIA [32]).

---

§ We note that the signal in Eq. (6) appears to scale as $\gamma^2$. However, this quadratic scaling is only an artefact arising from the definition of $\vec{B}_{ALP}$ which exhibit a $1/\gamma$ dependence. As such, the signal remain linear in $\gamma$.

In this procedure, the hyperpolarized $^{129}$Xe sample is immersed in a leading magnetic field, $\vec{B}_0$, oriented along the z-axis and the bulk magnetization is also initially along the z-axis. Prior to the acquisition, a $\pi/2$ magnetic pulse is applied to the sample. After the pulse, the magnetization is in the transverse plane: $M_z \xrightarrow{\pi/2} M_{xy}$. Under the action of $\vec{B}_0$, $\vec{M}_{xy}$ precesses about the z-axis at the Larmor frequency $\gamma|\vec{B}_0|$. During the precession, a transient signal $S(t)$ is acquired for a time $\tau = \min\{\tau_{ALP}, T_2\}$. Recalling that $\tau_{ALP} \sim 10^6/\omega_a$ and $T_2 \sim 10 - 1000$ s, then in this low-frequency regime $\tau_{ALP} > T_2$, and the transient-signal coherence time becomes $\tau = T_2$. Once the coherence time $T_2$ is reached, the transverse magnetization has decayed and the sample is switched for a new one. A $\pi/2$ pulse is applied again and the next transient acquisition takes place.

This measurement method differs from the previous one in the sense that it does not require the ALP wind to tilt the sample magnetization. Following a resonant $\pi/2$ pulse, the magnetization is always in the xy-plane, producing a signal oscillating at the Larmor frequency. The detection involves measuring modulation of the Larmor frequency, induced by the ALP-wind, similarly to the AC-Zeeman effect [42]. In contrast to the resonant scheme, sidebands are induced by the longitudinal component of $\vec{B}_{ALP}$ relative to $\vec{B}_0$: $\vec{B}_{ALP,z}$ (see Fig. 2-b). The frequency-modulated signal takes the form:

$$S(t) \propto \gamma\rho P e^{-i\gamma|\vec{B}_0|t} e^{-i\gamma|\vec{B}_{ALP,z}|\sin(\omega_a t)/\omega_a}. \tag{9}$$

This expression can be written in terms of the Bessel functions of the first kind $J_k$:

$$S(t) \propto \gamma\rho P \sum_{k=-\infty}^{\infty} J_k\left(\frac{\gamma|\vec{B}_{ALP,z}|}{\omega_a}\right) e^{-i\gamma|\vec{B}_0|t} e^{ik\omega_a t}. \tag{10}$$

The spectrum of such a frequency modulated signal exhibits a large central peak at the Larmor frequency and sidebands located at $\gamma|\vec{B}_0| \pm k\omega_a$, where $k = 1, 2, 3...$. The $k^{th}$ sideband's amplitude is given by [43]:

$$|S_k(t)| \propto \gamma\rho P\, J_k(I) \;\Rightarrow\; |S_k(t)| \propto \gamma\rho P\left(\frac{I^k}{2^k k!} - \frac{I^{k+2}}{2^{k+2}(k+1)!} + ...\right), \tag{11}$$

where $I = \gamma|\tilde{B}_{ALP,z}|/\omega_a$ is the modulation index. Recalling that $|\vec{B}_{ALP}| \sim 10^{-10}$–$10^{-30}$ T, we see that $I \ll 1$. Thus, the signal can be approximated by its carrier and the first set of sidebands ($k = \pm 1$) arising at frequencies $\gamma|\vec{B}_0| \pm \omega_a$ [44]. Expanding $J_1$ to first order, the sideband signal can be approximated by:

$$S_{SB}(t) \propto \gamma^2 \rho P \frac{|\vec{B}_{ALP,z}|}{\omega_a} e^{-i\gamma|\vec{B}_0|t} e^{\pm i\omega_a t}. \tag{12}$$

The Fourier transform of $S(t)$ yields a spectrum presenting a central peak at $\gamma|\vec{B}_0|$, surrounded by two sidebands located at $\gamma|\vec{B}_0| \pm \omega_a$. The amplitude of the central peak is large because the full magnetization is rotated by the $\pi/2$ pulse. As $\gamma|\vec{B}_{ALP,z}|/\omega_a \ll 1$, the sidebands amplitude is expected to be small.

Synchronizing the pulse with a stable clock enables recovery of initial phases of each transient signal and allows coherent averaging as long as the acquisition time is smaller than $\tau_{\text{ALP}}$ [45, 46]. Thus we can coherently average the transient signals into sets, $\bar{s}$, each of length $T_2$. Here $\bar{s}$, is the average over $n = \tau_{\text{ALP}}/T_2$ transient signals: $\bar{s} = \frac{1}{n}\sum_{i=1}^{n} S_i(t)$. Coherent averaging of $n$ transient signals decreases the expectation value of the white noise as $1/\sqrt{n}$ [36] while keeping the signal amplitude constant, as in the resonant case. Considering a white-noise spectral density of $\sqrt{S(f)}$, coherent averaging of the $n$ transient signals yields a signal threshold of:

$$A_{\text{min,n}} \sim \sqrt{S(f)}\,(nT_2)^{-1/2}. \quad (13)$$

We recall that the sidebands are always located around the Larmor frequency, therefore this scheme removes the need to search for resonance. $|\vec{B}_0|$ is arbitrarily adjusted such that the detection can be done in a region where the SQUID sensitivity is optimum, regardless of the ALP frequency: $\sqrt{S(f)} := \sqrt{S_{\text{opt}}}$. We can then rewrite Eq. (13) setting the SQUID noise spectral density to $\sqrt{S_{\text{opt}}}$:

$$A_{\text{min,n}} \sim \sqrt{S_{\text{opt}}}\,(nT_2)^{-1/2}. \quad (14)$$

The acquisition and coherent averaging are repeated until the desired total integration time $T_{\text{tot}}$ is reached. This yields $N$ independent averaged sets $\{\bar{s}_1, \bar{s}_2...\bar{s}_N\}$ each of length $T_2$ and of signal threshold $A_{\text{min,n}}$. Here $N = T_{\text{tot}}/\tau_{\text{ALP}}$, is the number of independent sets. Each set is measured for a time $\tau_{ALP}$. We recall that after $\tau_{\text{ALP}}$, the ALP-wind phase changes by an unknown amount. As a result, the $N$ sets have uncorrelated phases and averaging them would decrease the signal as well as the noise. However, some additional signal-processing techniques can be employed to improve the detection threshold, namely, incoherent averaging of the sets $\bar{s}_i$.

In practice, incoherent averaging corresponds to averaging the sets $\{\bar{s}_1, \bar{s}_2...\bar{s}_N\}$ in the frequency domain by averaging their power spectral densities: $\overline{\text{PSD}} = \frac{1}{N}\sum_{i=1}^{N}\text{PSD}(\bar{s}_i)$. Such processing does not reduce the noise mean value, but only the noise-power standard deviation [47]. After the incoherent averaging sequence, the signal threshold becomes (details are given in the appendix of Ref. [24]):

$$A_{\text{min,N}} \sim A_{\text{min,n}}\,N^{-1/4} \Rightarrow A_{\text{min,N}} \sim \sqrt{S_{\text{opt}}}\,(\tau_{\text{ALP}}\,T_{\text{tot}})^{-1/4}. \quad (15)$$

The signal threshold scales as $t^{-1/2}$ as long as the signal is phase coherent ($t < \tau_{\text{ALP}}$), then scales as $t^{-1/4}$ once the ALP coherence time is reached ($t > \tau_{\text{ALP}}$). Recalling that $\tau_{\text{ALP}} \sim 10^6/\omega_a$ the signal threshold in Eq. (15) becomes:

$$A_{\text{min,N}} \sim 10^{-3/2}\sqrt{S_{\text{opt}}}\,(T_{\text{tot}}\,\omega_a)^{-1/4}. \quad (16)$$

After the coherent and incoherent averaging sequences, the signal-to-noise ratio $\text{SNR}_{\text{sidebands}}$ is obtained by comparing the signal amplitude from Eq. (12) to $A_{\text{min,N}}$:

$$\text{SNR}_{\text{SB}} := |S_{\text{SB}}(t)|/A_{\text{min,N}} \quad (17)$$

$$\sim 10^{3/2}\gamma^2 \rho P |\vec{B}_{\text{ALP,z}}|\sqrt{S_{\text{opt}}}^{-1} T_{\text{tot}}^{1/4}\omega_a^{-5/4}. \quad (18)$$

$\text{SNR}_{\text{SB}}$ does not scale with $T_2$. Indeed, the transient signals can be coherently averaged until $\tau_{\text{ALP}}$ is reached, making $T_2$ irrelevant (ignoring the duty cycle). Two factors contribute to the gains in sensitivity in the low-frequency ALPs region; 1) Lower ALPs frequencies imply longer ALP-wind coherence time, increasing the time during which the signal can be coherently averaged 2) The amplitude of the transient signal is determined by the modulation index of the first Bessel function. As such, low frequency modulations produce higher sidebands amplitudes.

The relative sensitivity of the resonant and sideband schemes is determined by the ratio of equations (18) and (8). This ratio is computed by assuming identical samples and equal longitudinal and transverse components of $\vec{B}_{\text{ALP}}$. We impose identical total integration time, $T_{\text{tot}}$, for both the resonant and sidebands measurement scheme. $T_{\text{tot}}$ is calculated by assuming coherent averaging for a time $\tau = T_2$ in each frequency bin during the resonant search. Ignoring duty cycle, the total integration time during the resonant search becomes $T_{\text{tot}} = (|\Omega|/\delta\nu)T_2$. Where $\Omega$ is the frequency range of interest and $\delta\nu \sim 1/\pi T_2$ is the linewidth of the resonance. The relative sensitivity of the resonant and sideband scheme becomes:

$$\text{SNR}_{\text{SB}}/\text{SNR}_{\text{RES}} \sim 10^{3/2}|\Omega|^{1/4}\pi^{-1/4}\frac{\sqrt{S(f)}}{\sqrt{S_{\text{opt}}}}\,\omega_{\text{a}}^{-5/4}\,T_2^{-1}. \qquad (19)$$

We now set the frequency range to $\Omega$ = 0–1 kHz and assume that the SQUID white-noise level above the knee frequency, $f_0$ = 2 Hz, is approximately constant $\sqrt{S(f > f_0)} = \sqrt{S_{\text{opt}}} = 0.9$ fT/$\sqrt{\text{Hz}}$. Below $f_0$, the SQUID noise is set to $\sqrt{S(f < f_0)} \sim 10$ fT/$\sqrt{\text{Hz}}$ (extracted from Ref. [48]). Values of $T_2$ higher than 100 seconds impose a total integration time longer than 1 year to probe the region $\Omega$ via the resonant scheme and are ignored. Eq. (19) is plotted against $\omega_{\text{a}}$ and $T_2$ in Fig. 3.

As shown in Fig. 3, the sidebands scheme is beneficial for ALP frequencies below 80 Hz and if long resonant integration time in each frequency bin cannot be achieved (either due to low $T_2$ or to some constraint on $T_{\text{tot}}$). We recall that the sidebands are located around the central peak at $\gamma|\vec{B}_0| \pm \omega_{\text{a}}$. Thus if the ALP frequency is lower than the linewidth $1/\pi T_2$, the sidebands are inside the central peak and cannot be resolved. This case corresponds to the excluded, red region of Fig. 3 and represents the current lower bound of the CASPEr experiment. Considering a realistic $T_2$ of 100 seconds yields a lower limit in the mHz range. This sideband-detection scheme enables detection of ALPs with masses in the $m_{\text{a}} \sim 10^{-17}$–$10^{-14}$ eV region, increasing the bandwidth of the experiment by three orders of magnitude and allowing the CASPEr detection region to overlap with ultracold neutrons experiments [49].

## 4. Conclusion

Axions and ALPs are well-motivated dark-matter candidates; in addition, the QCD axion provides a solution to the strong CP problem. The discovery of such particles would shed light on many fundamental questions in modern physics, offering a glimpse of physics beyond the Standard Model.

The oscillatory pseudo-magnetic couplings between axion/ALPs and matter open the possibility of direct dark-matter detection via NMR techniques. When nuclear spins couple to ALPs in the Milky Way dark-matter halo, the spins behave as if they were in an oscillating magnetic field. The axion-gluon coupling can induce an oscillating nucleon electric-dipole moment. CASPEr-Wind and CASPEr-Electric seek to measure the NMR signals induced by the ALP wind and the nucleon EDM, respectively.

The original CASPEr experiment is based on resonant search via CW-NMR. This method enables the search for axions and ALPs at frequencies ranging from a few Hz to a few hundred MHz ($m_a \sim 10^{-14} - 10^{-6}$ eV). Probing lower frequencies using this approach suffers in sensitivity due to limitations of the magnetometers. Therefore non-resonant detection of ALP-induced sidebands around the Larmor frequency could be beneficial. This detection scheme allows probing of the axion and ALP parameter space in the mHz to Hz region ($m_a \sim 10^{-17}$–$10^{-14}$ eV), thus increasing the bandwidth of CASPEr by three decades. Sideband-based ALP searches introduced by the SILFIA experiment using hyperpolarized noble gases in the gas phase and SQUID detection [32], are already in progress at the The National Metrology Institute of Germany (PTB).

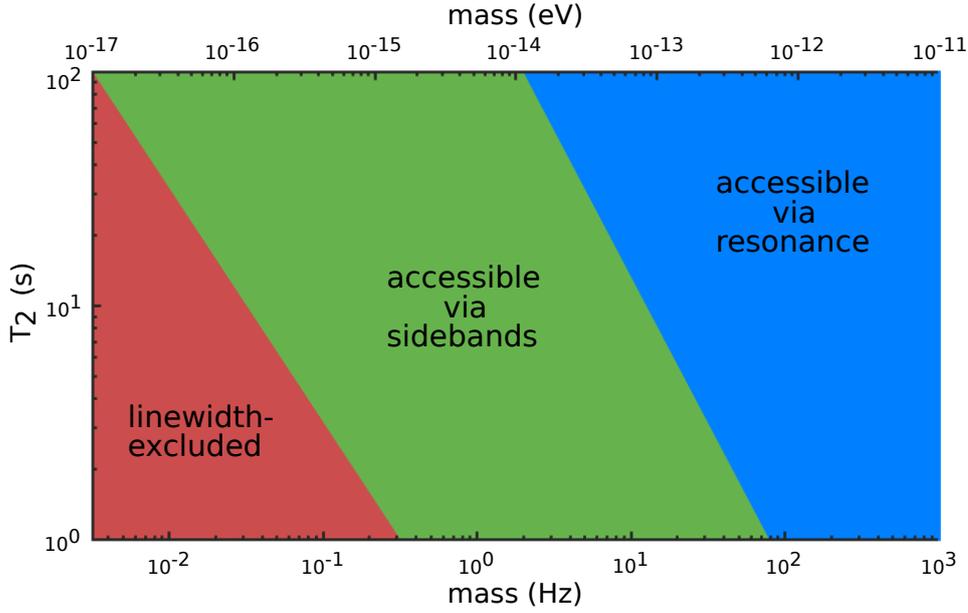

**Figure 3:** Resonant versus sidebands detection: sensitivity cut-off regions for equal total integration time and sample parameters (spin density, geometry). Extracted from Eq. (19). Blue area: sensitivity is higher for resonant detection. Green area: region enabled by sidebands detection. Red area: excluded by the linewidth $1/\pi T_2$. We assume that the SQUID white-noise level above the knee frequency, $f_0 = 2$ Hz, is approximately constant $\sqrt{S}(f > f_0) \sim 0.9$ fT/$\sqrt{\text{Hz}}$. Below $f_0$, SQUID noise is set to $\sqrt{S}(f < f_0) \sim 10$ fT/$\sqrt{\text{Hz}}$ (extracted from Ref. [48]; low-$T_c$ W9L-18D9 SQUID - PTB). Values of $T_2$ higher than 100 seconds impose a total integration time longer than 1 year to probe the region of interest via the resonant scheme and are ignored.

Only a few experiments are tuned to the mass range accessible by CASPEr, even though the presence of ALPs at these frequencies is well-motivated. This makes CASPEr complementary to other searches, which typically look at lower or higher frequencies. CASPEr-Wind and CASPEr-Electric are currently under construction and are scheduled to start acquiring data in early 2018.

**Acknowledgement:** The authors would like to thank Martin Engler, Anne Fabricant, Pavel Fadeev and Hector Masia Roig for useful discussions and comments. This project has received funding from the European Research Council (ERC) under the European Unions Horizon 2020 research and innovation programme (grant agreement No 695405). We acknowledge the support of the Simons and Heising-Simons Foundations and the DFG Reinhart Koselleck project.

# References

[1] E. Purcell, H. Torrey, and R. Pound. Resonance Absorption by Nuclear Magnetic Moments in a Solid. *Physical Review Letters*, 69(1-2):37–38, 1946.

[2] F. Bloch, W.W. Hansen, and M. Packard. The nuclear induction experiment. *Physical Review Letters*, 70(7-8):474–485, 1946.

[3] J. T. Arnold, S. S. Dharmatti, and M. E. Packard. Chemical effects on nuclear induction signals from organic compounds. *The Journal of Chemical Physics*, 19(4):507–507, 1951.

[4] M. Jiang, T. Wu, J.W. Blanchard, G. Feng, X. Peng, and D. Budker. Experimental benchmarking of quantum control in zero-field nuclear magnetic resonance. *arXiv:1708.06324v1*, 2017.

[5] J. Bian, M. Jiang, J. Cui, X. Liu, B. Chen, Y. Ji, B. Zhang, J.W. Blanchard, X. Peng, and J. Du. Universal quantum control in zero-field nuclear magnetic resonance. *Physical Review A*, 95, 2017.

[6] F. Zwicky. Die Rotverschiebung von extragalaktischen Nebeln. *Helvetica Physica Acta. 6: 110127*, 1933.

[7] C. Amsler et al. Review of Particle Physics. *Physics Letters, Section B: Nuclear, Elementary Particle and High-Energy Physics*, 667(1-5):1–6, 2008.

[8] R. D. Peccei and H.R. Quinn. CP conservation in the presence of pseudoparticles. *Physical Review Letters*, 38(25):1440–1443, 1977.

[9] S. Weinberg. A new light boson? *Physical Review Letters*, 40(4):223–226, 1978.

[10] F. Wilczek. Problem of Strong P and T Invariance in the Presence of Instantons. *Physical Review Letters*, 40(5), 1978.

[11] J. Ipser and P. Sikivie. Can Galactic Halos Be Made of Axions? *Physical Review Letters*, (12), 1983.

[12] L.F. Abbott and P. Sikivie. A cosmological bound on the invisible axion. *Physical Review Letters B*, 120, 1983.

[13] M. Dine and W. Fischler. The not so harmless axion. *Physical Review Letters B*, 120, 1983.

[14] J. Preskill, M.B. Wise, and F. Wilczek. Cosmology of the invisible axion. *Physical Review Letters B*, 120(1-3):127–132, 1983.

[15] D.J.E. Marsh. Axion cosmology. *Physics Reports*, 643(2):1–79, 2016.

[16] M. Srednicki. Axion couplings to matter. (I). CP-conserving parts. *Nuclear Physics, Section B*, 260(3-4):689–700, 1985.

[17] H. Primakoff. Photo-Production of Neutral Mesons in Nuclear Electric Fields and the Mean Life of the Neutral Meson. *Physical Review Letters*, 802(243), 1950.

[18] K. Zioutas et al. First results from the cern axion solar telescope. *Physical Review Letters*, 94:121301, 2005.

[19] G. Raffelt and L. Stodolsky. Mixing of the photon with low-mass particles. *Physical Review Letters D*, 37(5), 1988.

[20] K. Ehret et al. Resonant laser power build-up in ALPS-A "light shining through a wall" experiment. *Nuclear Instruments and Methods in Physics Research, Section A: Accelerators, Spectrometers, Detectors and Associated Equipment*, 612(1):83–96, 2009.

[21] S.J. Asztalos et al. SQUID-based microwave cavity search for dark-matter axions. *Physical Review Letters*, 104(4):1–4, 2010.

[22] P. Sikivie. Experimental Tests of the Invisible Axion. *Physical Review Letters*, 51(16), 1983.

[23] P.W. Graham, I.G. Irastorza, S.K. Lamoreaux, A. Lindner, and K.A. Van Bibber. Experimental Searches for the Axion and Axion-Like Particles. *Annual Review of Nuclear and Particle Science*, 65(1):485–514, 2015.

[24] D. Budker, P.W. Graham, M. Ledbetter, S.Rajendran, and A.O. Sushkov. Proposal for a cosmic axion spin precession experiment (CASPEr). *Physical Review Letters X*, 4(2):1–10, 2014.

[25] P.W. Graham and S. Rajendran. New observables for direct detection of axion dark matter. *Physical Review Letters D - Particles, Fields, Gravitation and Cosmology*, 88(3):1–16, 2013.

[26] R. Catena and P. Ullio. A novel determination of the local dark matter density. *arXiv:0907.0018v2*, 2009.

[27] G.G. Raffelt. *Astrophysical Axion Bounds*, pages 51–71. Springer Berlin Heidelberg, Berlin, Heidelberg, 2008.

[28] M.I. Vysotsky, Y.B. Zeldovich, M.Y. Khlopov, and V.M. Chechetkin. Some astrophysical limitations on the axion mass. *Journal of Experimental and Theoretical Physics Letters*, 27, 1978.

[29] P.W. Graham, I.G. Irastorza, S.K. Lamoreaux, A. Lindner, and K.A. Van Bibber. Experimental Searches for the Axion and Axion-Like Particles. 2015.

[30] M.H. Levitt. *Spin Dynamics: Basics of Nuclear Magnetic Resonance - 2nd Ed, ch.2–3, p.34–39,*. Wiley, 2008.

[31] A. Derevianko. Detecting dark matter waves with precision measurement tools. *arXiv:1605.09717v2*, 2016.

[32] L. Trahms, private communication.

[33] P.W. Graham and S. Rajendran. Axion dark matter detection with cold molecules. *Physical Review Letters D*, 84, 2011.

[34] J. Jaeckel. A force beyond the Standard Model Status of the quest for hidden photons. *Frascati Physics Series*, 56, 2012.

[35] S. Chaudhuri, P.W. Graham, K.Irwin, J. Mardon, S. Rajendran, and Y. Zhao. A radio for hidden-photon dark matter detection. *Physical Review Letters D*, 92, 2015.

[36] J.H. John. *An Introduction to Digital Signal Processing, Ch.11, p.258*. Academic Press, 1991.

[37] M.P. Ledbetter and M.V.Romalis. Nonlinear Effects from Dipolar Interactions in Hyperpolarized Liquid 129 Xe. *Physical Review Letters*, 89.

[38] M.V. Romalis and M.P. Ledbetter. Transverse Spin Relaxation in Liquid 129 Xe in the Presence of Large Dipolar Fields. *Physical Review Letters*, (33).

[39] B. Driehuys, G.D. Cates, E. Miron, K. Sauer, D.K. Walter, and W. Happer. High-volume production of laser-polarized 129 Xe. *Applied Physics Letters*, 69, 1996.

[40] T. Wang, D.F.J. Kimball, A.O. Sushkov, D. Aybas, J.W. Blanchard, G. Centers, S.R.O. Kelley, J. Fang, and D. Budker. Application of Spin-Exchange Relaxation-Free Magnetometry to the Cosmic Axion Spin Precession Experiment. *arXiv:1701.08082*, 2017.

[41] D. Robbes. Highly sensitive magnetometers-a review. *Sensors and Actuators, A: Physical*, 129(1-2 SPEC. ISS.):86–93, 2006.

[42] D. Budker, D. Kimball, and D. DeMille. *Atomic physics: An exploration through problems and solutions, 2nd Ed. , p.90,95,109,121*. 2008.

[43] G. Arfken. *Mathematical Methods for Physicists, 6th Ed., Ch.11, p.675–693*, volume 40. 1972.

[44] D. Budker and S. Rocherster. *Optically Polarized Atoms, Ch.4, p.75–81*.

[45] J. M. Boss, K. S. Cujia, J. Zopes, and C. L. Degen. Quantum sensing with arbitrary frequency resolution. *Science*, 356(6340):837–840, 2017.

[46] D.B. Bucher, D.R. Glenn, J. Lee, M.D. Lukin, and H. Park. High resolution magnetic resonance spectroscopy using solid-state spins. *arXiv:1705.08887v2*, 2017.

[47] R.G. Lyons. *Understanding Digital Signal Processing, Ch.8, p.320–330*, volume 40. 2004.

[48] D. Drung. High-performance DC SQUID read-out electronics. *Physica C: Superconductivity and its Applications*, 368(1-4):134–140, 2002.

[49] C. Abel et al. Search for axion-like dark matter through nuclear spin precession in electric and magnetic fields. *arXiv:1708.06367*, 2017.